\documentclass[aip, jap, reprint]{revtex4-1}
\usepackage{amsmath,amssymb}
\usepackage{graphicx}\usepackage{multirow}

\begin{document}

\title{Influence of Cooling on Quantum Network Deployment}%


\author{Tobias Schaich}
\email[]{tcs49@cam.ac.uk}

\author{Vasileios Karavias}
\affiliation{University of Cambridge, Cambridge, UK}

\author{Max Sich}

\author{Scott Dufferwiel}
\affiliation{AegiQ, Sheffield, UK}

\author{Mike Payne}
\affiliation{University of Cambridge, Cambridge, UK}%

\author{Andrew Lord}
\affiliation{BT, Adastral Park, Ipswich, UK}

\date{\today}
  \begin{abstract}
    Superconducting detectors for quantum key distribution require expensive cooling but have advantageous detection properties. A linear programming algorithm and heuristic for network cooling placement are proposed and validated. Co-locating cooled detectors provides cost-effective solutions even when cooling costs exceed the equipment cost of multiple links.
    \end{abstract}

\maketitle                  

\section{Introduction}

The promise of secure communication based on Quantum Key Distribution (QKD) has resulted in a steady increase of its technological maturity since the first quantum key exchange protocol was proposed in 1984 \cite{Bennett2014}. Advanced protocols such as Decoy State BB84 have demonstrated increased security and reach \cite{Hwang2003, Zhao2006} and commercial QKD equipment for use in optical networks has reached the market \cite{Quantique2020}. But, QKD networks are still subject to active research\cite{Elliott2007, Alleaume2014, Sasaki2011, Qiu2014, Zhang2018}. A recent overview of major QKD networks points out that current implementations suffer from detector dark counts constraining the QKD channel length \cite{Mehic2020}. 

A detector solution with comparatively few dark counts are superconducting nanowires\cite{Hadfield2009, Gemmell2017}. However, these operate at extremely low temperatures, requiring expensive cryostats. Thus, the idea of co-locating single photon detectors at a single network node such that cooling resources are shared may hold merit for reducing overall deployment cost. Fundamentally, this changes how QKD is deployed in an existing optical network.

Sharing resources in this manner has been overlooked in studies on QKD equipment placement in optical networks \cite{Alleaume2009, Li2020, WANG2020, Pederzolli2020}. Consequently, we here establish the benefits of co-locating cooling in a QKD network. We propose a linear programming algorithm and heuristic for cooling placement in a given network and investigate the cost threshold for cooled solutions to become economically favourable to uncooled solutions. 

\section{QKD Link Capacity}

QKD key rates are modelled for the Decoy State BB84 protocol. We use a Poissonian source, an optical fibre and photon detectors whose relevant parameters are defined and tabulated in Tab.~\ref{tab:variables}. For a length $l$ of fibre, define the fibre efficiency as $\eta_f = 10^{-\frac{\alpha l}{10}}$. A 3.5~ns detector window is assumed giving a dark count probability of $p_{DC}=r_{DC}\times 3.5~\text{ns}$. For n-photon yield $Y_n = [1-(1-\eta_f\eta_d)^n] + [1-\eta_f\eta_d]^n p_{DC}$, the probability of a detection event given a source trigger event is:\cite{Lo2005, Fung2006} 
\begin{equation*}
    p_{\mu} = \sum_{n=0} \frac{e^{-\mu}\mu^n}{n!}  Y_n.
\end{equation*}
The fraction of states not lost due to the detector dead time is \cite{Eraerds2010} $\eta_{dead} = (1+\tau_{dead}f_{rep}p_{\mu})^{-1}$ resulting in a system gain $Q_{\mu} = p_{\mu} \eta_{dead}.$

\begin{figure}[t]
    \centering
    \includegraphics[width=0.43\textwidth]{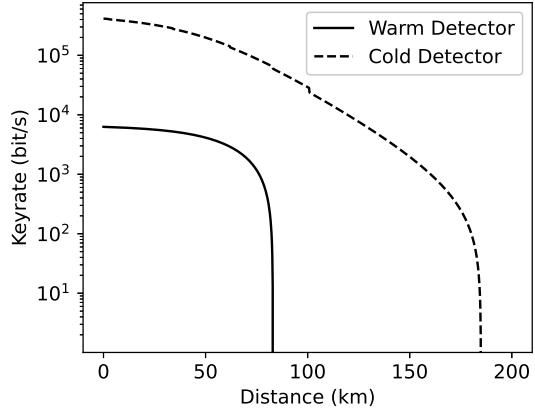}
    \caption{Quantum secure key rates calculated using Eq. \eqref{eq:keyrate}.}
    \label{fig:Capacities}
\end{figure}

For BB84 the Quantum Bit Error Rate (QBER) is\cite{Gobby2004}  $E_{\mu} = (1-V)/2$ for a visibility $V$ approximated by 
\begin{equation*}
    V = \frac{\mu \eta_f\eta_{d}}{\mu \eta_f \eta_{d} + 2P_{e}},
\end{equation*}
with $P_e$  modelled as $P_e = (5.3 \times 10^{-7} + p_{DC})$\cite{Gobby2004}. Simultaneously the QBER is given by: 
\begin{equation*}
    E_{\mu} = \frac{\eta_{dead}}{Q_\mu}(\sum_{n=0} Y_n \frac{e^{-\mu}\mu^n}{n!}e_n),
\end{equation*}
which allows fitting of the n-photon state QBER $e_n$\cite{Lo2005}. The resulting secure key rate is \cite{Lo2005} 
\begin{equation}
    R = -Q_{\mu}f(E_{\mu})H_2(E_{\mu}) + Q_1[1-H_{2}(e_1)],
    \label{eq:keyrate}
\end{equation}
with single photon gain $Q_{1} = Y_1\mu e^{-\mu} \eta_{dead}$, Shannon entropy $H_2$ and error correcting code efficiency $f(E_{\mu}) = 1.2$ in our model\cite{Eraerds2010}. Keyrate over distance is presented in Fig. \ref{fig:Capacities} showing that indeed cooled detectors produce higher keyrates and larger reach than uncooled solutions. 

\begin{table}[t]
    \centering
    \begin{tabular}{|c|c|}
        \hline
        Mean photon number $\mu$ & 0.1 \\
        \hline
        Fibre loss $\alpha$ & 0.2~dB/km\\
        \hline
        Pulse repetition rate $f_{rep}$ & 100~MHz\\
        \hline
        \multirow{2}{*}{Dark count rate $r_{DC}$ \cite{Quantique2021a, Quantique2021}}  & 100~Hz (cold)\\
         & 6~kHz (warm)\\
        \hline
        \multirow{2}{*}{Detector efficiency $\eta_d$ \cite{Quantique2021a, Quantique2021}}  & 0.85 (cold)\\
         & 0.2 (warm)\\
        \hline
        \multirow{2}{*}{Dead time $\tau_{dead}$ \cite{Quantique2021a, Quantique2021}} & 1~$\mu$s (cold)\\
         & 50~$\mu$s (warm)\\ \hline
    \end{tabular}
    \caption{Photon source, fibre and photon detector properties.}
    \label{tab:variables}
\end{table}

\section{Graph Topology}
To assess the usefulness of cooling in a network, we define the following model. A network is represented by a directed graph $g=(N,E)$ consisting of a set of communicating trusted nodes $N$ and edges $E$ representing optical channels. Nodes are randomly placed in a $100 \times 100~\text{km}$ box. Edges are placed by successively connecting each node to its nearest neighbour if no connection already exists. Then, each node is connected to their second to nearest neighbour etc. When a critical average node degree is reached, no further connections are placed. 

As long links can have low or zero capacity, we introduce additional nodes that bisect an edge if it cannot support a minimum key rate in the uncooled case. These nodes, understood as trusted relays, require transmission and receiving equipment which increases overall cost. Hence, the cost-minimising algorithm introduced in the next section can omit equipping a relay.

\section{Cost Minimising Algorithm}

Given a graph $g$, we implement a linear programming algorithm to find the cheapest solution which can support a demand matrix $\mathbf{K}$ whose elements $K_{ij}$ represent the key material which node $i$ must exchange with node $j$. The hop-by-hop keys exchange is modelled as a multi-commodity flow \cite{Orlin1994, Alleaume2014, Pederzolli2020} with variables $x_{nm}^{(i,j)}$ describing the amount of key material flowing along the edge $(n,m)$ with source node $i$ and destination node $j$. The set of all source-destination pairs is $\mathcal{K}$. Each edge has a capacity $c_{nm}^{warm}$ and $c_{nm}^{cool}$ given by Eq.~\eqref{eq:keyrate}. Furthermore, we introduce integer variables $\delta_{nm}$ indicating the number of links equipped with communications equipment and binary variables $\xi_n$ which represent if a node is cooled. Only cost associated with equipping links with QKD equipment and cooling are considered (no operational cost). Cost per link is normalised to unity and the cooling cost relative to the link cost is denoted as $C_C$. The complete linear program is:
\begin{align}
\begin{split}
    &\text{minimise}:~    \sum_{(n,m) \in E} \delta_{nm} + \sum_{n \in N} \xi_n C_C\\
	&\text{subject to}: \hfill \\
    &\sum_{(i,j) \in \mathcal{K}} (x^{(i,j)}_{nm})-\delta_{nm} c^{cool}_{nm} \leq 0 \\
    &\sum_{(i,j) \in \mathcal{K}}(x^{(i,j)}_{nm}-\xi_m K_{ij})-\delta_{nm}c^{warm}_{nm}\leq 0 \\
    &\sum_{n} (x^{(i,j)}_{in})+\sum_{n} (x_{ni}^{(j,i)})=K_{ij}+K_{ji} \\
    &\sum_{m}(x_{nm}^{(i,j)})-\sum_{m}(x_{mn}^{(i,j)})=0, ~ \forall n\in N \backslash \{i,j\}   \\
    &x^{(i,j)}_{ni} = 0, ~ \forall n \in N \text{ with } (n,i) \in E \\
    &x^{(i,j)}_{jn} = 0,  ~ \forall n \in N \text{ with } (j,n) \in E \\
    &~\xi_n \in \{0,1\},~\delta_{nm} \in \mathbb{N}_0,~x^{(i,j)}_{nm} \in \mathbb{N}_0  
\end{split}
\label{eq:optimising_algo}
\end{align}
Unless otherwise stated, each line is valid for all $(i,j) \in \mathcal{K}$ and $(n,m) \in E$. The first two constraints combined ensure that the correct link capacity is utilised whereas the other constraints ensure key flow conservation \cite{Orlin1994}. Source-destination pair $(i,j)$ and $(j,i)$ are treated symmetrically mirroring the symmetric key cipher \cite{Pederzolli2020}.

\textbf{Cooled Nodes Placement Heuristic.} Further to the presented optimal algorithm, we propose a heuristic for placing cooling in a given network. The idea is that cooling will benefit the most links if it is placed at the node with the most connections. Hence, the heuristic strategy sequentially places cooling at the nodes with the highest degree for a given number of cooled nodes. Then, the algorithm in \eqref{eq:optimising_algo} with fixed $\xi_n$ is invoked. This method converges faster than the algorithm with dynamic cooling placement. The optimal topology for a given cooling cost is determined in post-processing by finding the minimum of the cost-function over the topologies with any number of cooled nodes. 

\section{Results and Discussion}
For our simulations, we use IBM Cplex\cite{Cplex2020} to solve \eqref{eq:optimising_algo} on graphs with mean node degree of 3.5. Trusted relays are deployed such that every link can support at least 4~kbit/s. We assume a full traffic matrix with $K_{ij}=8/(|N|-1)~\text{kbit/s}$ where $|N|$ is the number of communicating nodes (excl. trusted relays). So, each trusted node introduces $8~\text{kbit/s}$ of traffic to the network, indicating a linear traffic growth with $|N|$. For simplicity, we exclusively considered single link networks letting $\delta_{nm}$ be binary. Only graphs which could support the demand matrix without cooling under this constraint were analyzed limiting  $|N|$ to 10. 

\begin{figure}
    \centering
    \includegraphics[width=0.41\textwidth]{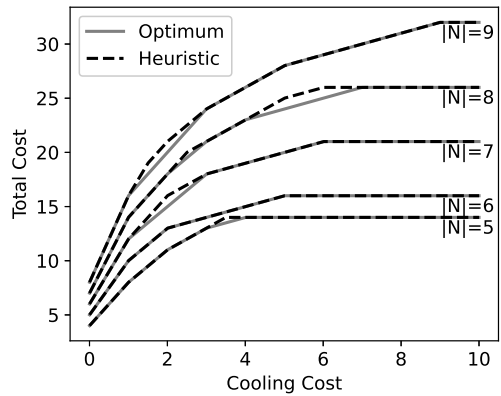}
    \caption{Heuristic and optimal network cost as a function of cooling cost $C_C$ for a graph with $|N|$ communicating nodes.}
    \label{fig:OptimalVsHeuristic}
\end{figure}

Figure \ref{fig:OptimalVsHeuristic} shows the calculated total cost against the cooling cost for graphs containing 5 to 9 communicating nodes. Focusing on the solid lines, their gradient corresponds to the number of cooled nodes in the optimal network. As a different topology becomes optimal, the gradient changes. Thus, high cooling costs induce flat curves indicating no cooled nodes. Equivalently, for cheap cooling the gradient is greatest as a large number of nodes is cooled. Fig. \ref{fig:OptimalVsHeuristic} also shows how the heuristic strategy compares against the optimal placement. Mostly, the heuristic solution agrees with the optimum. However, some deviations from the optimum are observed, indicating that the placement heuristic is not always optimal. Sub-optimal performance may occur if the graph has a low-degree node requiring a lot of trusted relays. Despite that, the heuristic performs well within 10\% of the optimal solution.

Following our validation efforts, we investigated the \emph{critical cooling cost} for which cooling a single node is cheaper than a completely uncooled solution. It corresponds to the number of QKD links which do not need to be equipped due to cooling. So, this metric is crucial to the economic viability analysis of future QKD network providers.

\begin{figure}
     \centering
     \includegraphics[width=0.43\textwidth]{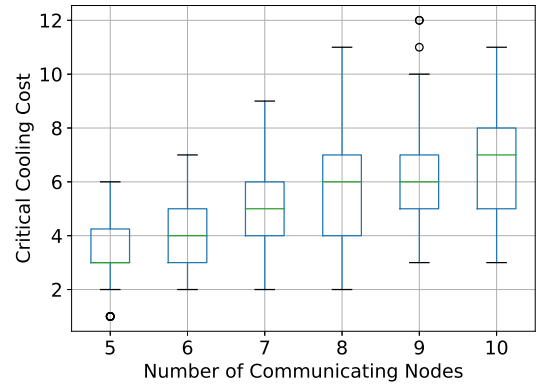}
     \caption{Box plot showing the critical cooling cost for which cooling a single trusted node reduces overall cost.}
     \label{fig:Boxplot}
 \end{figure}

A statistical approach was chosen to investigate the critical cooling cost. Using the heuristic method, it was calculated for 100 instances of graphs having 5-10 communicating nodes. The resulting data, shown in the box plot in Fig. \ref{fig:Boxplot}, indicate that the median critical cooling cost grows from 3 to 7 with the number of communicating nodes. This correlates with the increased traffic which more communicating nodes introduce to the network. Similarly, the minimum critical cost shifts from 1 to 3, the maximum from 6 to 11 for more communicating nodes. 

We estimate current cryostat cost at about 0.6-0.7 in our units. In addition, cooled detectors are about 5$\times$ more expensive than uncooled detectors which are an estimated 5\% of an uncooled link's cost. For a single cooled node, we can combine these costs to an effective cooling cost $C_C$ which lies between 1 and 2 reaching the upper bound with increased number of co-located detectors. Even in the worst case, these results show that median savings up to 5 link costs can be expected which can increase to 9-10 link costs in certain topologies.  Further economic benefits of co-location of cryogenically cooled detectors can be expected in the future as the cooling technology matures, and prices reduce.

\section{Conclusions and Outlook}

We have proposed an optimal and heuristic placement algorithm for applying cooling in a QKD network. The heuristic provides good results within less than 10\% of the optimum solution. Co-locating shared cooling equipment is then considered from an economic perspective. We provide evidence that a cooled solution can provide significant savings in the current technology landscape. Our data suggest that the cost benefit increases with the number of communicating nodes and network traffic. An optimisation including cooled, high-performance sources offering extended link reach, presents an interesting research opportunity as further link savings are expected.

\section{Acknowledgements}
We thank The Royal Society (INF-PHD-180021), EPSRC (EP/V519662/1) and BT for their support.


%

\begin{thebibliography}{24}%
	\makeatletter
	\providecommand \@ifxundefined [1]{%
		\@ifx{#1\undefined}
	}%
	\providecommand \@ifnum [1]{%
		\ifnum #1\expandafter \@firstoftwo
		\else \expandafter \@secondoftwo
		\fi
	}%
	\providecommand \@ifx [1]{%
		\ifx #1\expandafter \@firstoftwo
		\else \expandafter \@secondoftwo
		\fi
	}%
	\providecommand \natexlab [1]{#1}%
	\providecommand \enquote  [1]{``#1''}%
	\providecommand \bibnamefont  [1]{#1}%
	\providecommand \bibfnamefont [1]{#1}%
	\providecommand \citenamefont [1]{#1}%
	\providecommand \href@noop [0]{\@secondoftwo}%
	\providecommand \href [0]{\begingroup \@sanitize@url \@href}%
	\providecommand \@href[1]{\@@startlink{#1}\@@href}%
	\providecommand \@@href[1]{\endgroup#1\@@endlink}%
	\providecommand \@sanitize@url [0]{\catcode `\\12\catcode `\$12\catcode
		`\&12\catcode `\#12\catcode `\^12\catcode `\_12\catcode `\%12\relax}%
	\providecommand \@@startlink[1]{}%
	\providecommand \@@endlink[0]{}%
	\providecommand \url  [0]{\begingroup\@sanitize@url \@url }%
	\providecommand \@url [1]{\endgroup\@href {#1}{\urlprefix }}%
	\providecommand \urlprefix  [0]{URL }%
	\providecommand \Eprint [0]{\href }%
	\providecommand \doibase [0]{http://dx.doi.org/}%
	\providecommand \selectlanguage [0]{\@gobble}%
	\providecommand \bibinfo  [0]{\@secondoftwo}%
	\providecommand \bibfield  [0]{\@secondoftwo}%
	\providecommand \translation [1]{[#1]}%
	\providecommand \BibitemOpen [0]{}%
	\providecommand \bibitemStop [0]{}%
	\providecommand \bibitemNoStop [0]{.\EOS\space}%
	\providecommand \EOS [0]{\spacefactor3000\relax}%
	\providecommand \BibitemShut  [1]{\csname bibitem#1\endcsname}%
	\let\auto@bib@innerbib\@empty
	\bibitem [{\citenamefont {Bennett}\ and\ \citenamefont
		{Brassard}(2014)}]{Bennett2014}%
	\BibitemOpen
	\bibfield  {author} {\bibinfo {author} {\bibfnamefont {C.~H.}\ \bibnamefont
			{Bennett}}\ and\ \bibinfo {author} {\bibfnamefont {G.}~\bibnamefont
			{Brassard}},\ }\href@noop {} {\bibfield  {journal} {\bibinfo  {journal}
			{Theoretical Computer Science}\ }\textbf {\bibinfo {volume} {560}},\ \bibinfo
		{pages} {7} (\bibinfo {year} {2014})}\BibitemShut {NoStop}%
	\bibitem [{\citenamefont {Hwang}(2003)}]{Hwang2003}%
	\BibitemOpen
	\bibfield  {author} {\bibinfo {author} {\bibfnamefont {W.~Y.}\ \bibnamefont
			{Hwang}},\ }\href@noop {} {\bibfield  {journal} {\bibinfo  {journal}
			{Physical Review Letters}\ }\textbf {\bibinfo {volume} {91}},\ \bibinfo
		{pages} {057901} (\bibinfo {year} {2003})}\BibitemShut {NoStop}%
	\bibitem [{\citenamefont {Zhao}\ \emph {et~al.}(2006)\citenamefont {Zhao},
		\citenamefont {Qi}, \citenamefont {Ma}, \citenamefont {Lo},\ and\
		\citenamefont {Qian}}]{Zhao2006}%
	\BibitemOpen
	\bibfield  {author} {\bibinfo {author} {\bibfnamefont {Y.}~\bibnamefont
			{Zhao}}, \bibinfo {author} {\bibfnamefont {B.}~\bibnamefont {Qi}}, \bibinfo
		{author} {\bibfnamefont {X.}~\bibnamefont {Ma}}, \bibinfo {author}
		{\bibfnamefont {H.~K.}\ \bibnamefont {Lo}}, \ and\ \bibinfo {author}
		{\bibfnamefont {L.}~\bibnamefont {Qian}},\ }\href@noop {} {\bibfield
		{journal} {\bibinfo  {journal} {Physical Review Letters}\ }\textbf {\bibinfo
			{volume} {96}},\ \bibinfo {pages} {070502} (\bibinfo {year} {2006})}\BibitemShut
	{NoStop}%
	\bibitem [{\citenamefont {Quantique}(2020)}]{Quantique2020}%
	\BibitemOpen
	\bibfield  {author} {\bibinfo {author} {\bibfnamefont {I.}~\bibnamefont
			{Quantique}},\ }\href
	{https://marketing.idquantique.com/acton/attachment/11868/f-021a/1/-/-/-/-/Cerberis
		QKD System{\_}Brochure.pdf} {\enquote {\bibinfo {title} {{Cerberis3 QKD
					System - Product Brochure}},}\ } (\bibinfo {year} {2020})\BibitemShut
	{NoStop}%
	\bibitem [{\citenamefont {Elliott}\ and\ \citenamefont
		{Yeh}(2007)}]{Elliott2007}%
	\BibitemOpen
	\bibfield  {author} {\bibinfo {author} {\bibfnamefont {C.}~\bibnamefont
			{Elliott}}\ and\ \bibinfo {author} {\bibfnamefont {H.}~\bibnamefont {Yeh}},\
	}\href@noop {} {\enquote {\bibinfo {title} {{DARPA Quantum Network
					Testbed}},}\ }\bibinfo {type} {Tech. Rep.}\ (\bibinfo  {institution} {BBN
		Technologies},\ \bibinfo {address} {Cambridge, Massachusetts},\ \bibinfo
	{year} {2007})\BibitemShut {NoStop}%
	\bibitem [{\citenamefont {All{\'{e}}aume}\ \emph {et~al.}(2014)\citenamefont
		{All{\'{e}}aume} \emph {et~al.}}]{Alleaume2014}%
	\BibitemOpen
	\bibfield  {author} {\bibinfo {author} {\bibfnamefont {R.}~\bibnamefont
			{All{\'{e}}aume}} \emph {et~al.},\ }\href@noop {} {\bibfield  {journal}
		{\bibinfo  {journal} {Theoretical Computer Science}\ }\textbf {\bibinfo
			{volume} {560}},\ \bibinfo {pages} {62} (\bibinfo {year} {2014})}\BibitemShut
	{NoStop}%
	\bibitem [{\citenamefont {Sasaki}\ \emph {et~al.}(2011)\citenamefont {Sasaki}
		\emph {et~al.}}]{Sasaki2011}%
	\BibitemOpen
	\bibfield  {author} {\bibinfo {author} {\bibfnamefont {M.}~\bibnamefont
			{Sasaki}} \emph {et~al.},\ }\href@noop {} {\bibfield  {journal} {\bibinfo
			{journal} {Optics Express}\ }\textbf {\bibinfo {volume} {19}},\ \bibinfo
		{pages} {10387} (\bibinfo {year} {2011})}\BibitemShut {NoStop}%
	\bibitem [{\citenamefont {Qiu}(2014)}]{Qiu2014}%
	\BibitemOpen
	\bibfield  {author} {\bibinfo {author} {\bibfnamefont {J.}~\bibnamefont
			{Qiu}},\ }\href@noop {} {\bibfield  {journal} {\bibinfo  {journal} {Nature}\
		}\textbf {\bibinfo {volume} {508}},\ \bibinfo {pages} {441} (\bibinfo {year}
		{2014})}\BibitemShut {NoStop}%
	\bibitem [{\citenamefont {Zhang}\ \emph {et~al.}(2018)\citenamefont {Zhang},
		\citenamefont {Xu}, \citenamefont {Chen}, \citenamefont {Peng},\ and\
		\citenamefont {Pan}}]{Zhang2018}%
	\BibitemOpen
	\bibfield  {author} {\bibinfo {author} {\bibfnamefont {Q.}~\bibnamefont
			{Zhang}}, \bibinfo {author} {\bibfnamefont {F.}~\bibnamefont {Xu}}, \bibinfo
		{author} {\bibfnamefont {Y.-A.}\ \bibnamefont {Chen}}, \bibinfo {author}
		{\bibfnamefont {C.-Z.}\ \bibnamefont {Peng}}, \ and\ \bibinfo {author}
		{\bibfnamefont {J.-W.}\ \bibnamefont {Pan}},\ }\href@noop {} {\bibfield
		{journal} {\bibinfo  {journal} {Optics Express}\ }\textbf {\bibinfo {volume}
			{26}},\ \bibinfo {pages} {24260} (\bibinfo {year} {2018})}\BibitemShut
	{NoStop}%
	\bibitem [{\citenamefont {Mehic}\ \emph {et~al.}(2020)\citenamefont {Mehic}
		\emph {et~al.}}]{Mehic2020}%
	\BibitemOpen
	\bibfield  {author} {\bibinfo {author} {\bibfnamefont {M.}~\bibnamefont
			{Mehic}} \emph {et~al.},\ }\href@noop {} {\bibfield  {journal} {\bibinfo
			{journal} {ACM Computing Surveys}\ }\textbf {\bibinfo {volume} {53}},\ \bibinfo {pages} {96}
		(\bibinfo {year} {2020})}\BibitemShut {NoStop}%
	\bibitem [{\citenamefont {Hadfield}(2009)}]{Hadfield2009}%
	\BibitemOpen
	\bibfield  {author} {\bibinfo {author} {\bibfnamefont {R.~H.}\ \bibnamefont
			{Hadfield}},\ }\href@noop {} {\bibfield  {journal} {\bibinfo  {journal}
			{Nature Photonics}\ }\textbf {\bibinfo {volume} {3}},\ \bibinfo {pages} {696}
		(\bibinfo {year} {2009})}\BibitemShut {NoStop}%
	\bibitem [{\citenamefont {Gemmell}\ \emph {et~al.}(2017)\citenamefont {Gemmell}
		\emph {et~al.}}]{Gemmell2017}%
	\BibitemOpen
	\bibfield  {author} {\bibinfo {author} {\bibfnamefont {N.~R.}\ \bibnamefont
			{Gemmell}} \emph {et~al.},\ }\href@noop {} {\bibfield  {journal} {\bibinfo
			{journal} {Superconductor Science and Technology}\ }\textbf {\bibinfo
			{volume} {30}},\ \bibinfo
		{pages} {11LT01} (\bibinfo {year} {2017})}\BibitemShut {NoStop}%
	\bibitem [{\citenamefont {All{\'{e}}aume}\ \emph {et~al.}(2009)\citenamefont
		{All{\'{e}}aume}, \citenamefont {Roueff}, \citenamefont {Diamanti},\ and\
		\citenamefont {L{\"{u}}tkenhaus}}]{Alleaume2009}%
	\BibitemOpen
	\bibfield  {author} {\bibinfo {author} {\bibfnamefont {R.}~\bibnamefont
			{All{\'{e}}aume}}, \bibinfo {author} {\bibfnamefont {F.}~\bibnamefont
			{Roueff}}, \bibinfo {author} {\bibfnamefont {E.}~\bibnamefont {Diamanti}}, \
		and\ \bibinfo {author} {\bibfnamefont {N.}~\bibnamefont {L{\"{u}}tkenhaus}},\
	}\href@noop {} {\bibfield  {journal} {\bibinfo  {journal} {New Journal of
				Physics}\ }\textbf {\bibinfo {volume} {11}} ,\ \bibinfo
			{pages} {075002}  (\bibinfo {year}
		{2009})}\BibitemShut {NoStop}%
	\bibitem [{\citenamefont {Li}\ \emph {et~al.}(2020)\citenamefont {Li},
		\citenamefont {Wang}, \citenamefont {Mao}, \citenamefont {Yao},\ and\
		\citenamefont {Han}}]{Li2020}%
	\BibitemOpen
	\bibfield  {author} {\bibinfo {author} {\bibfnamefont {Q.}~\bibnamefont
			{Li}}, \bibinfo {author} {\bibfnamefont {Y.}~\bibnamefont {Wang}}, \bibinfo
		{author} {\bibfnamefont {H.}~\bibnamefont {Mao}}, \bibinfo {author}
		{\bibfnamefont {J.}~\bibnamefont {Yao}}, \ and\ \bibinfo {author}
		{\bibfnamefont {Q.}~\bibnamefont {Han}},\ }\href@noop {} {\bibfield
		{journal} {\bibinfo  {journal} {Optics Express}\ }\textbf {\bibinfo {volume}
			{28}},\ \bibinfo {pages} {9419} (\bibinfo {year} {2020})}\BibitemShut
	{NoStop}%
	\bibitem [{\citenamefont {Wang}\ \emph {et~al.}(2020)\citenamefont {Wang},
		\citenamefont {Li}, \citenamefont {Mao}, \citenamefont {Han}, \citenamefont
		{Huang},\ and\ \citenamefont {Xu}}]{WANG2020}%
	\BibitemOpen
	\bibfield  {author} {\bibinfo {author} {\bibfnamefont {Y.}~\bibnamefont
			{Wang}}, \bibinfo {author} {\bibfnamefont {Q.}~\bibnamefont {Li}}, \bibinfo
		{author} {\bibfnamefont {H.}~\bibnamefont {Mao}}, \bibinfo {author}
		{\bibfnamefont {Q.}~\bibnamefont {Han}}, \bibinfo {author} {\bibfnamefont
			{F.}~\bibnamefont {Huang}}, \ and\ \bibinfo {author} {\bibfnamefont
			{H.}~\bibnamefont {Xu}},\ }\href@noop {} {\bibfield  {journal} {\bibinfo
			{journal} {Optics Express}\ }\textbf {\bibinfo {volume} {28}},\ \bibinfo
		{pages} {26348} (\bibinfo {year} {2020})}\BibitemShut {NoStop}%
	\bibitem [{\citenamefont {Pederzolli}, \citenamefont {Faticanti},\ and\
		\citenamefont {Siracusa}(2020)}]{Pederzolli2020}%
	\BibitemOpen
	\bibfield  {author} {\bibinfo {author} {\bibfnamefont {F.}~\bibnamefont
			{Pederzolli}}, \bibinfo {author} {\bibfnamefont {F.}~\bibnamefont
			{Faticanti}}, \ and\ \bibinfo {author} {\bibfnamefont {D.}~\bibnamefont
			{Siracusa}},\ }\href@noop {} {\bibfield  {journal} {\bibinfo  {journal}
			{Quantum Reports}\ }\textbf {\bibinfo {volume} {2}},\ \bibinfo {pages} {114}
		(\bibinfo {year} {2020})}\BibitemShut {NoStop}%
	\bibitem [{\citenamefont {Lo}, \citenamefont {Ma},\ and\ \citenamefont
		{Chen}(2005)}]{Lo2005}%
	\BibitemOpen
	\bibfield  {author} {\bibinfo {author} {\bibfnamefont {H.~K.}\ \bibnamefont
			{Lo}}, \bibinfo {author} {\bibfnamefont {X.}~\bibnamefont {Ma}}, \ and\
		\bibinfo {author} {\bibfnamefont {K.}~\bibnamefont {Chen}},\ }\href@noop {}
	{\bibfield  {journal} {\bibinfo  {journal} {Physical Review Letters}\
		}\textbf {\bibinfo {volume} {94}},\ \bibinfo {pages} {230504} (\bibinfo {year}
		{2005})}\BibitemShut {NoStop}%
	\bibitem [{\citenamefont {Fung}, \citenamefont {Tamaki},\ and\ \citenamefont
		{Lo}(2006)}]{Fung2006}%
	\BibitemOpen
	\bibfield  {author} {\bibinfo {author} {\bibfnamefont {C.~H.~F.}\
			\bibnamefont {Fung}}, \bibinfo {author} {\bibfnamefont {K.}~\bibnamefont
			{Tamaki}}, \ and\ \bibinfo {author} {\bibfnamefont {H.~K.}\ \bibnamefont
			{Lo}},\ }\href@noop {} {\bibfield  {journal} {\bibinfo  {journal} {Physical
				Review A - Atomic, Molecular, and Optical Physics}\ }\textbf {\bibinfo
			{volume} {73}},\ \bibinfo {pages} {012337} (\bibinfo {year} {2006})}\BibitemShut
	{NoStop}%
	\bibitem [{\citenamefont {Eraerds}\ \emph {et~al.}(2010)\citenamefont
		{Eraerds}, \citenamefont {Walenta}, \citenamefont {Legr{\'{e}}},
		\citenamefont {Gisin},\ and\ \citenamefont {Zbinden}}]{Eraerds2010}%
	\BibitemOpen
	\bibfield  {author} {\bibinfo {author} {\bibfnamefont {P.}~\bibnamefont
			{Eraerds}}, \bibinfo {author} {\bibfnamefont {N.}~\bibnamefont {Walenta}},
		\bibinfo {author} {\bibfnamefont {M.}~\bibnamefont {Legr{\'{e}}}}, \bibinfo
		{author} {\bibfnamefont {N.}~\bibnamefont {Gisin}}, \ and\ \bibinfo {author}
		{\bibfnamefont {H.}~\bibnamefont {Zbinden}},\ }\href@noop {} {\bibfield
		{journal} {\bibinfo  {journal} {New Journal of Physics}\ }\textbf {\bibinfo
			{volume} {12}},\ \bibinfo {pages} {063027}  (\bibinfo {year} {2010})}\BibitemShut {NoStop}%
	\bibitem [{\citenamefont {Gobby}, \citenamefont {Yuan},\ and\ \citenamefont
		{Shields}(2004)}]{Gobby2004}%
	\BibitemOpen
	\bibfield  {author} {\bibinfo {author} {\bibfnamefont {C.}~\bibnamefont
			{Gobby}}, \bibinfo {author} {\bibfnamefont {Z.~L.}\ \bibnamefont {Yuan}}, \
		and\ \bibinfo {author} {\bibfnamefont {A.~J.}\ \bibnamefont {Shields}},\
	}\href@noop {} {\bibfield  {journal} {\bibinfo  {journal} {Applied Physics
				Letters}\ }\textbf {\bibinfo {volume} {84}},\ \bibinfo {pages} {3762}
		(\bibinfo {year} {2004})}\BibitemShut {NoStop}%
	\bibitem [{\citenamefont {Quantique}(2021{\natexlab{a}})}]{Quantique2021a}%
	\BibitemOpen
	\bibfield  {author} {\bibinfo {author} {\bibfnamefont {I.}~\bibnamefont
			{Quantique}},\ }\href
	{https://marketing.idquantique.com/acton/attachment/11868/f-89626e7c-0af7-4151-b267-c3ff356656d7/1/-/-/-/-/ID
		Qube NIR Free-Running Brochure.pdf} {\enquote {\bibinfo {title} {{ID Qube NIR
					Free-running - Product Brochure}},}\ } (\bibinfo {year}
	{2021}{\natexlab{a}})\BibitemShut {NoStop}%
	\bibitem [{\citenamefont {Quantique}(2021{\natexlab{b}})}]{Quantique2021}%
	\BibitemOpen
	\bibfield  {author} {\bibinfo {author} {\bibfnamefont {I.}~\bibnamefont
			{Quantique}},\ }\href
	{https://marketing.idquantique.com/acton/attachment/11868/f-023b/1/-/-/-/-/ID281{\_}Brochure.pdf}
	{\enquote {\bibinfo {title} {{ID281 Superconducting nanowire system - Product
					Brochure}},}\ } (\bibinfo {year} {2021}{\natexlab{b}})\BibitemShut {NoStop}%
	\bibitem [{\citenamefont {Ravindra}, \citenamefont {Magnati},\ and\
		\citenamefont {Orlin}(1993)}]{Orlin1994}%
	\BibitemOpen
	\bibfield  {author} {\bibinfo {author} {\bibfnamefont {A.}~\bibnamefont
			{Ravindra}}, \bibinfo {author} {\bibfnamefont {T.}~\bibnamefont {Magnati}}, \
		and\ \bibinfo {author} {\bibfnamefont {J.}~\bibnamefont {Orlin}},\
	}\href@noop {} {\emph {\bibinfo {title} {{Network flows: Theory, algorithms
					and applications}}}}\ (\bibinfo  {publisher} {Prentice Hall Inc.},\ \bibinfo
	{address} {New Jersey},\ \bibinfo {year} {1993})\BibitemShut {NoStop}%
	\bibitem [{\citenamefont {IBM}(2020)}]{Cplex2020}%
	\BibitemOpen
	\bibfield  {author} {\bibinfo {author} {\bibnamefont {IBM}},\ }\href@noop {}
	{\enquote {\bibinfo {title} {{IBM ILOG CPLEX 12.10 User's Manual}},}\ }
	(\bibinfo {year} {2020})\BibitemShut {NoStop}%
\end{thebibliography}

%


\end{document}